\documentclass[Letter, 10 pt, conference]{ieeeconf}
\usepackage[utf8]{inputenc}
\usepackage[T1]{fontenc}
\usepackage{amsmath}
\usepackage{amssymb}
\usepackage{amsfonts}
\usepackage{mathtools}
\usepackage{mathtools}
\usepackage{mathrsfs}
\usepackage{graphicx}
\usepackage{caption}
\usepackage{babel}
\usepackage{balance}
\usepackage{cite}
\usepackage{dsfont}
\usepackage{graphicx}
\usepackage{floatrow}
\usepackage{xcolor}

\newcommand{\probP}{\text{I\kern-0.15em P}}

\IEEEoverridecommandlockouts

\let\theoremstyle\relax
\usepackage{amsthm}

\usepackage{float}
\floatstyle{plaintop}
\restylefloat{table}

\theoremstyle{definition}

\newtheorem{assumption}{Assumption}

\newtheorem{problem}{Problem}
\newtheorem{remark}{Remark}

\usepackage{comment}
\usepackage{todonotes}

\title{Conference2022}
\title{\LARGE \bf On Safety of Passengers Entering a Bus Rapid Transit System from Scheduled Stops}

\author{Alejandra Valencia, {\itshape{Student Member, IEEE}}, Andreas A. Malikopoulos, {\itshape{Senior Member, IEEE}}
\thanks{This work was supported by NSF under Grants CNS-2149520 and CMMI-2219761.}
\thanks{The authors are with the Department of Mechanical Engineering, University of Delaware, Newark, DE 19716 USA (emails: \tt\small{aleval@udel.edu}; \tt\small{andreas@udel.edu}.)}}

\date{March 2022}

\begin{document}

\maketitle

\begin{abstract} 
In this paper, we address the vehicle scheduling problem for improving passenger safety in bus rapid transit systems. Our focus is on passengers waiting at street stops to enter terminal stations. To enhance their safety, we minimize deviations from the proposed timetable, thereby minimizing passengers' initial waiting time. We formulate an optimization problem considering the position, speed deviation, and passenger count at each stop, solved using dynamic programming. Numerical simulations validate the effectiveness of our approach in enhancing passenger safety. Our work is the first attempt to minimize waiting time for improved safety and the first to utilize position tracking for departure time matching.
\end{abstract}

\section{Introduction}
\label{sec:Intro}
The growth to enhance the safety demand for energy-efficient and safe mobility is driving significant transformations in social transportation practices \cite{chalaki2021CSM,Chremos2020SocialDilemma}. Furthermore, with advancements in transportation technology, there is an expectation of improved efficiency, safety, and reduced travel time, thereby addressing congestion issues \cite{Zhao2018ITSC,malikopoulos2019ACC,Malikopoulos2020}. In the context of urban mobility, public transportation plays a vital role. Still, it also presents challenges for passengers regarding route planning and choosing the appropriate travel modes \cite{bast2016route,zhao2019enhanced}.
The global journey planning problem is closely intertwined with optimizing urban mobility resources and their interactions \cite{zhang2021unveiling}. Traditional public transportation planning focuses on creating line plans, constructing timetables, and allocating resources to ensure customer satisfaction and efficient service execution \cite{van2018vehicle}. However, these static scheduling somehow disregard the dynamic characteristics of transportation networks \cite{d2019dynamic}. Recent research focuses on optimization strategies considering passenger safety and real-time control for public transport synchronization \cite{gkiotsalitis2022review}. These approaches aim to minimize passenger travel times \cite{li2018integrated, bie2020dynamic, moosavi2020using}, waiting times \cite{hadas2010optimal, sadrani2022vehicle, hartmann2020optimization}, or logistics costs \cite{sun2019flexible} while coordinating holding and synchronized transfers \cite{delgado2013holding}. However, some optimizations neglect passenger convenience, focusing solely on timetabling or vehicle scheduling \cite{bouman2020new, ren2021optimal}, which is a critical aspect.

Passenger safety perception in public transport networks has also been extensively studied. Factors such as crime indicators and speed limits influence passenger safety perception \cite{ceccato2022crime, pranoto2016improvement}. Transit nodes, including bus stops, can act as crime absorbers, with crowdedness at bus system nodes leading to crimes such as fights, pick-pocketing, and sexual harassment. On the other hand, empty transit environments increase the likelihood of violent assaults. Minimizing the time spent at bus stops can reduce passengers' exposure to unsafe environments \cite{tucker2003safer}, positively impacting their safety. Gender also plays a role, as many women feel unsafe at bus stops at night due to the fear of sexual harassment \cite{ceccato2022fear, fan2016waiting}. In cities with higher levels of harassment, only 1-7\% of female users feel safe waiting at stops, while in cities with low levels, 67-88\% feel safe \cite{ceccato2022fear}. During different stages of a transportation trip, safety perceptions can vary. Integrated systems like bus rapid transit (BRT) can increase anxiety, particularly for female users, if waiting times are prolonged \cite{fan2016waiting}, which falls out the overall passenger convenience.

Our study addresses passenger safety in a BRT system by optimizing on-time bus performance using a given timetable as a baseline. We utilize real-time information to minimize the number of passengers waiting at each stop. This optimization approach incorporates tracking the position, speed, and passenger count at each time instant, allowing us to penalize deviations from the timetable, speed variations, and exposure to unsafe environments.  To the best of our knowledge, including position tracking for on-time performance is a novel contribution to the literature. Enhancing passenger safety at bus stops is a primary objective of our work. By addressing passenger safety concerns and optimizing BRT performance, our study aims to improve public transportation systems, ensuring a safer and more efficient travel experience for passengers.

The remainder of the paper proceeds as follows. In Section \ref{sec:ModelingFramework}, we provide the modeling framework for maximizing the safety of the passengers by at least following the timetable and some constraints. In Section \ref{sec:control_framework}, we present the solution approach to solve the previous control problem. In Section \ref{sec:Simulation}, we provide a detailed analysis and simulation results. Finally, we draw concluding remarks in Section \ref{sec:conclusion}.

\section{Vehicle scheduling problem in a BRT system environment} \label{sec:ModelingFramework}
\begin{figure}[ht!]
    \includegraphics[trim={3cm 3cm 3cm 3cm},clip, width=0.99 \linewidth]{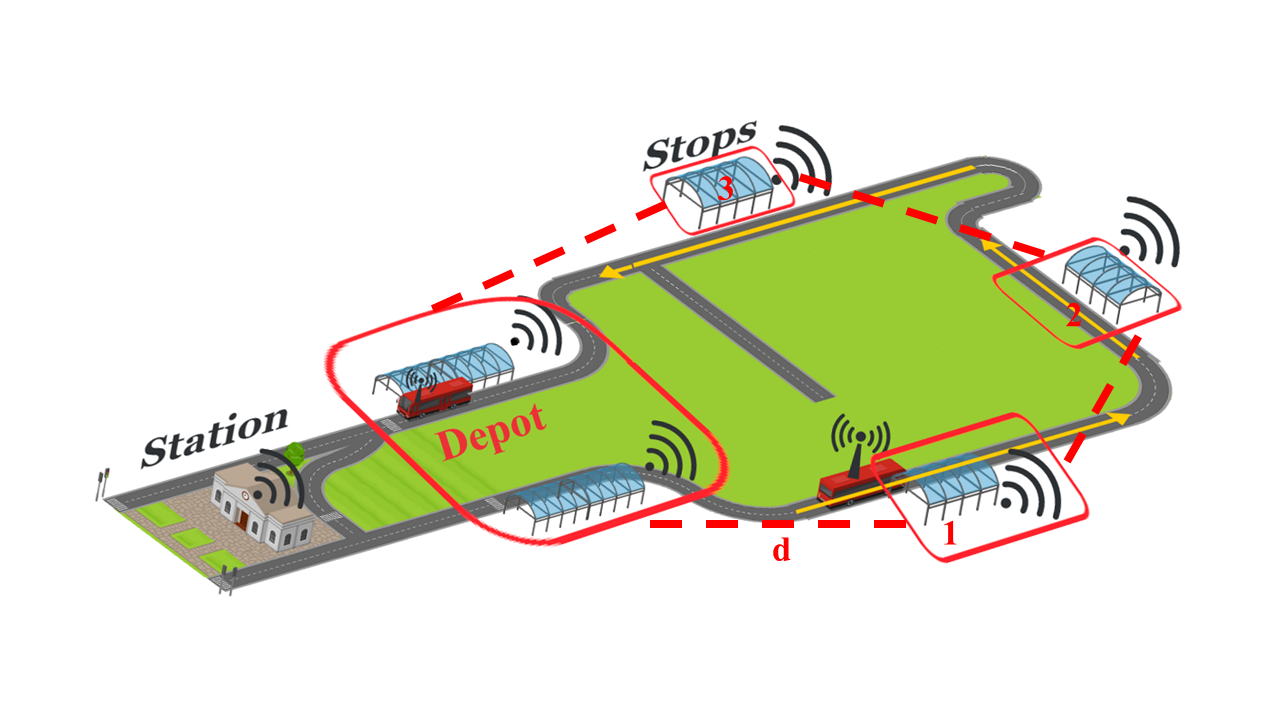}
    \caption{Example of a single loop with simple depot network.}
    \label{fig:singleloop}
\end{figure}

This study focuses on a Bus Rapid Transit (BRT) system with buses serving regular sidewalk bus stops in a defined environment. The system operates on a single loop with a single depot, as shown in Fig. \ref{fig:singleloop}. We aim to address the \emph{single depot vehicle scheduling problem} (SDVSP) to maximize passenger safety at terminal stations. We specifically consider passengers waiting to board at bus stops, excluding transfers between stations. Passengers are categorized as planning or non-planning, based on their flexibility in arrival time \cite{ansari2021waiting}. Planning passengers have specific desired departure times and adjust their schedules accordingly \cite{nuzzolo2012schedule}, while non-planning passengers arrive randomly \cite{de2012information}. This differentiation is significant as non-planning passengers often experience longer waiting times and potential safety risks on the street.


\subsection{Notation}
We use subscripts to denote time and superscripts to denote set indexes. Random variables are represented by uppercase letters, while their realizations are denoted by lowercase letters. For instance, $X_k$ represents the random variable, and $x_k$ represents its realization. The expected value of a random variable is denoted by $\mathbb{E}(\cdot)$, and the probability of an event is denoted by $\probP(\cdot)$. Feasible sets are denoted by uppercase cursive letters. Specifically, $\mathcal{H}$ represents the set of buses, $\mathcal{M}$ denotes the set of stops, $\mathcal{U}$ the set control strategy values, $\mathcal{G}$ the set of arriving passengers, and $\mathcal{A}$ the set of alighting passengers. Constants are denoted by lowercase cursive letters without subscripts. For example, $c^h$ represents the traffic factor associated with each bus, $\delta^h$ is the flexibility factor, and $\theta^m$ is the fixed location of the stops. Time stages are denoted by $k$, the control action is represented by $U_k$, and the system's state is denoted by $X_k$. The system state includes the position $P_k$, the most recent stop $M_k$, the available bus capacity $B_k$, and the list of passengers at each stop $N_k(m_k)$, where $m_k$ represents the current stop. The uncertain nature of the system is denoted by $W_k$, which encompasses the alighting passengers $A_k(m_k)$ and arriving passengers $G_k(m_k)$ at each stop.

\subsection{Problem Formulation}
In our problem formulation, we consider a single loop network with $h \in \mathbb{N}$ buses and passengers with flexible arrival times. The departure time of the passengers is fixed, and their arrival at a stop follows a random distribution that can vary during peak hours. The network is represented by a directed graph $\mathscr{G} = (\mathcal{M}, \mathcal{L})$, where $\mathcal{M}$ is the set of stops and $\mathcal{L}$ is the set of links connecting those stops. Specifically, we focus on a single loop network with a single depot, as depicted in Fig. \ref{fig:singleloop}.

To model the system's dynamics, we define discrete time stages denoted by $k$, each representing one second. The state of the system at time $k$ is denoted by $X_k$, and the control action at the same time is denoted by $U_k$. The set of possible control actions is $\mathcal{U} = \{0, 1, ..., U\}$, representing the speed of the bus. To address SDVSP, we consider a given timetable (also called \emph{tt}) that specifies each stop's arrival, waiting, and departure times. This timetable also associates a minimum desired position within the loop for each time instant. The position of the bus concerning the origin can be estimated using the random variable $P_k$. We denote the set of buses as $\mathcal{H} = \{1, \dots, H\}$, where $H$ represents the total number of buses considered in the analysis. The position of each bus evolves over time according to \eqref{eq:pos_ev}, where $c^h$ is a constant specific to each bus $h\in\mathcal{H}$, representing the distance traveled by the bus at a speed of $u_k = 1$ per time instant $k$. This constant takes into account the traffic conditions on the road. Therefore, the evolution of the position can be expressed as
\begin{equation}
P_{k+1} = P_{k} + U_k \cdotp c^h
\label{eq:pos_ev}.
\end{equation}
\begin{remark}
The minimum desired position inside the loop ensures that the bus waits at least a certain amount of time at each stop and always departs on time. The focus is not on controlling the exact arrival time at the stop.
\end{remark}
\begin{remark}
    The constant $c^h$, referred to as the \emph{traffic factor} considers the random nature of traffic in the loop. This modeling approach is appropriate for a BRT system where buses may travel on regular streets or dedicated bus lanes. The traffic factor captures the stochasticity of road congestion and is specific to each bus, varying at the start of each trip based on various conditions.
\end{remark}
The stops in the loop are positioned at specific locations, as shown in Fig. \ref{fig:singleloop}, where the positions of the stops are measured relative to the depot (indicated by the dashed line). In this example, stop $1$ is located at position $d$, and stop $2$ is located at position $2d$. We use the notation $\mathcal{M}=\{0,1,\dots, M\}$ to represent the set of stops, where $M \in\mathbb{N}$ denotes the total number of stops considered in our analysis. The location of each stop is denoted by ${\theta}^{m_k}$, where $m_k$ corresponds to the specific stop. To account for flexibility in detecting the stops, we introduce the factor $\delta^h$, which is associated with the bus $h$. This factor allows the bus to identify the stop within a range of positions rather than requiring a precise match.
\begin{figure}[ht!]
    \centering
    \includegraphics[trim={0 0 0 3.2cm},clip, width=0.69 \linewidth]{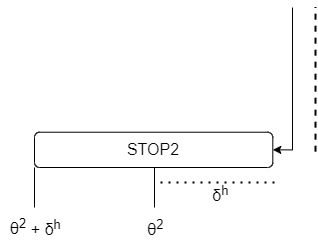}
    \caption{Relationship of ${\theta}^{m_k}$ and $\delta^h$.}
    \label{fig:delta}
\end{figure}

Fig \ref{fig:delta} graphically explains the relationship between ${\theta}^{m_k}$ and $\delta^h$, where in stop 2, $m_k$= 2, and ${\theta}^{m_k}$ = ${\theta}^{2}$. The stop starts for the bus at ${\theta}^{m_k}$ - $\delta^h$, and ends at ${\theta}^{m_k}$ + $\delta^h$. For this purpose, ${\theta}^0$ represents the position of the depot, which is the origin and destination of the route.  The most recent stop visited by the bus up to time $k$ is denoted by the random variable $M_k \in \mathcal{M}$,
\begin{equation}
M_{k+1} =  
\begin{cases}
    M_k + 1 & \text{ if  } \mathds{1} \{{\theta}^{m_k+1}-\delta^h\leq P_k< \\&\text{ }\theta^{m_k+1}+\delta^h\} \\&\text{ and  } u_k = 0\\
    M_k &\text{ otherwise}
\end{cases},
\label{eq:stops}
\end{equation}
 where $\mathds{1}\{arg\}$ is an indicator function, which returns 1 when the argument is true.
\begin{remark}
To address the possibility of buses missing stops due to traffic factors ($c^h$), we introduce the flexibility factor ($\delta^h$) for each bus. By allowing a range of positions for each stop instead of a fixed position, we account for the variability in bus movements. The value of $\delta^h$ is determined based on $c^h$ and determines the extent of the range within which the bus can detect and initiate the stop. This approach ensures the bus can effectively identify and stop at the designated location.
\end{remark}
The bus capacity is dynamically updated at each bus stop based on passenger boarding and alighting. We define the random variable ${N}_k({m_k})$ to represent the number of passengers waiting at the most recent stop, which takes values from the finite set $\mathcal{N}$. Similarly, ${A}_k({m_k})$ represents the number of passengers alighting at the current stop at time $k$, and it takes values from the finite set $\mathcal{A}$. At each instant, $l$ passengers can board the bus while at a stop. The binary flag $\gamma$ indicates whether a passenger intends to alight at the next stop. The random variable $B_k$ represents the available bus capacity, which takes values from the finite set $\mathcal{B}$.
\begin{equation}
B_{k+1} = 
\begin{cases}
    B_k & \text{if  } U_k > 0\\
    B_k - l + \gamma{A}_k({m_k})  & \text{if  } {N}_k({m_k}) > l\\
    B_k - {N}_k({m_k}) + \gamma{A}_k({m_k}) & \text{if  } {N}_k({m_k}) \leq l\\
\end{cases},
\label{eq:available}
\end{equation}
The number of passengers at the most recent stop, at each instant of time $k$, is denoted by ${N}_k({m_k})$, where if $m_k = {M_{k-1}}+1 \text{ and } U_k = 0$, which means the bus just stopped, and the stop was just updated, then
\small
\begin{equation}
    {N}_{k+1}({m_k})={N}_k({m_k})+{G}_{k-1}({m_k})+ 
    (B_{k+1}-B_k-\gamma{{A_k}(m_k)})
\label{eq:pass},
\end{equation}
\normalsize
otherwise, we have
\begin{equation}
{N}_{k+1}({m_k}) = 
    {N}_k({m_k})+{G}_{k-1}({m_k})
\label{eq:pass1},
\end{equation}
where ${G}_{k-1}({m_k})$ takes values in the set $\mathcal{G}$ and represents the distribution of the new passengers arriving at the stop at each time step. This variable and ${A}_k({m_k})$, are stochastic and will be considered as lognormal distributions, according to the data used in \cite{filabadi2022new}. All of these variables are part of the state of the system, which is represented by $X_k = (P_k, M_k, B_k, {N}_k({m_k}))$. Starting with the initial state $X_0$, the state evolves as $X_{k+1} = f_k (X_k, U_k, W_k)$. The random variable $W_k$ contains the uncertain nature of the formulation, and it is defined as $W_k = ({A}_k({m_k}), {G}_k({m_k}))$. These two variables are independent of each other. The control action $U_k$ is constrained. It takes values in the set $\mathcal{U} = \{0,0.5,1,1.5...,U\}$ where $U$ represents the maximum speed allowed. Then, $u_k \in \mathcal{U}$ and $\lambda$ is a limit factor in ensuring smooth speed changes. For the bus in between the stops, the feasible set is given by $U_{k+1} \neq 0 \text{ and } U_{k+1} \leq U \text{ and } \\U_{k+1} \in \{u_k\cdotp c - \lambda \cdotp c, u_k, u_k\cdotp c + \lambda \cdotp c\}$, where
\begin{equation}
\begin{cases}
     &\text{ if } \displaystyle\sum_{m=1} ^{M}\mathds{1} \{{\theta}^{m}+\delta^h \leq P_k < {\theta}^{{m}+1}-\delta^h\}>0\\&\text{ or } (B_k \leq 0 \text{ and } \gamma =0)
\end{cases}
\label{eq:concons},
\end{equation}
when the bus is approaching a stop, the set is given by, 
\begin{equation}
    U_{k+1} \in \{\max (u_k\cdotp c - \lambda \cdotp c, \lambda \cdotp c)\} \\\text{ if } P_k \leq \displaystyle\sum_{u=0}  ^{U} c \cdotp u 
    \label{eq:decel}.
\end{equation}
For the bus to stop, several conditions need to be satisfied, and they are specified in \eqref{eq:stoping}
\begin{equation}
U_{k+1} = 0
    \begin{cases}
      &\text{ if } \{N_k(m_k) > 0 \text{ or } \gamma =1\} \\&\text{and} \displaystyle\sum_{m=1}^{M}\mathds{1}\{{\theta}^{m}-\delta^h \leq P_k< {\theta}^{m}+\delta^h\}>0 
    \end{cases}
    \label{eq:stoping},
\end{equation}
which considers the passengers waiting at the stops, the willingness to alight of passengers in the bus, and the position of the bus with respect to the fixed stop as illustrated in Fig. \ref{fig:delta}. Finally, for the bus to depart from the stop, at least 3 of these conditions need to be satisfied, where the bus has to be at a stop, it should be waiting, and either no more passengers are waiting, the bus is full, or the timetable indicates departure. The feasible set is then just one value, given by $U_{k+1} \neq 0 \text{ and } U_{k+1} = \{u_k\cdotp c + \lambda \cdotp c\}$, where
\begin{equation}
    \begin{cases}
      &\text{ if } \displaystyle\sum_{m=1} ^{M}\mathds{1} \{{\theta}^{m}-\delta^h \leq P_k < {\theta}^{m}+\delta^h\}>0 \\&\text{ and } u_k = 0 \\&\text{ and } \{N_k(m_k) = 0 \text{ or } b_k \leq 0\} \\&\text{ and} (k - tt(k)) \leq l\} \\&\text{ or } (k - tt(k)) = 0\}
    \end{cases}
    \label{eq:depart},
\end{equation}
where the timetable is always a hard constraint for the bus to \emph{depart} from the stop \emph{on-time}, and it can only speed up by the limit factor $\lambda$.
With \eqref{eq:concons}-\eqref{eq:depart}, the feasible values of the control set $\forall \text{ } k$ are defined.
\begin{remark}
The right-hand side of condition \eqref{eq:decel} incorporates a summation over the entire control set, multiplied by the traffic factor. This calculation ensures that the bus can decelerate gradually within the specified limit factor $\lambda$. Considering the entire control set, the bus is given enough time to decelerate while maintaining clear visibility of the stop and promoting safety until it reaches a complete stop.
\end{remark}
We formulate the optimization problem as a minimization task to prioritize passenger safety. The objective is to minimize the waiting time of all passengers while considering a penalty for speed deviations. To achieve an \emph{on-time} performance with respect to the timetable, we aim to arrive at each stop as soon as possible without departing before the indicated time. We design the cost function \eqref{eq:J}, which incorporates the number of passengers at each stop, the bus position, and speed. According to the timetable, the desired position is represented by $pd^{h}_k$, while the actual position is denoted by $P^{h}_k$. In addition, we consider a speed range $ud \subset \mathcal{U}$, denoted by $ud = \{c,...,\mathcal{U}(U-1)\}$. The variable $d$ is 0 when the bus is at the depot and 1 elsewhere. For safety considerations, as mentioned in Section \ref{sec:Intro}, we aim to minimize the number of passengers at each stop and the speed deviation from the specified range. The speed range encompasses all speeds from the traffic factor matching speed to the second highest speed. Thus, any allowed speed that deviates from the desired range (either too slow or fast) is penalized. This penalty is addressed by incorporating the quadratic difference between each bus's current speed $U^{h}_k$ and the min and max speeds within the range.
Deviation from the speed range is considered undesirable. Traveling below the range may fail to meet the timetable, while exceeding the range could increase the risk of accidents and compromise passenger safety. The cost function includes weight factors $\alpha_1$, $\alpha_2$, and $\alpha_3$ to assign appropriate weights to different cost components. So, the cost is expressed as
  \begin{equation}
  \begin{aligned}
J_k =   \displaystyle\sum_{m=1} ^{M} \alpha_1 {N}_{k}(m) +\displaystyle\sum_{h=1} ^{H} \biggl\{\alpha_2\cdotp\big({{{pd}^{h}_{k} - {P}^{h}_{k}}}\big) + \\ d\cdotp\alpha_3\biggl(\big({U}^{h}_{k} - \min({ud}^{h})\big)^2 +\big({U}^{h}_{k} - \max({ud}^{h})\big)^2\biggl)\biggl\},
 \end{aligned}
 \label{eq:J}
\end{equation}
which represents a multi-objective function. Finally, the optimization problem is formulated in Problem 1.
\begin{problem}
The formulation of the optimization problem is as follows:
        \begin{gather}
            \min \displaystyle \mathop{\mathbb{E}}_{{\textbf{G}_{k}}}\Big\{J_k(x_k)  + \{J_{k+1}(x_{k+1})| x_{k}, u_{k}\}\Big\} \\
            \text{subject to:} \notag \\
            \eqref{eq:pos_ev} , \eqref{eq:stops}, \eqref{eq:available} , \eqref{eq:pass} , \eqref{eq:pass1} , \eqref{eq:concons}, \eqref{eq:decel} , \eqref{eq:stoping} , \eqref{eq:depart} \notag\\
            \text{and} \notag \\
            {A}_k({M_k}) \leq  B_{k-1} \label{eq:ali},
        \end{gather}
    where constraints \eqref{eq:pos_ev} to \eqref{eq:pass1} represent the evolution of the state, \eqref{eq:concons} to \eqref{eq:depart} the control constraints, and constraint \eqref{eq:ali} says that the number of passengers alighting the bus at a stop has to be less or equal than the passengers traveling on the bus.
    \label{problem1}
\end{problem}
\section{Dynamic solution approach} \label{sec:control_framework}
\begin{figure}[ht!]
    \centering
    \includegraphics[width=0.89\linewidth]{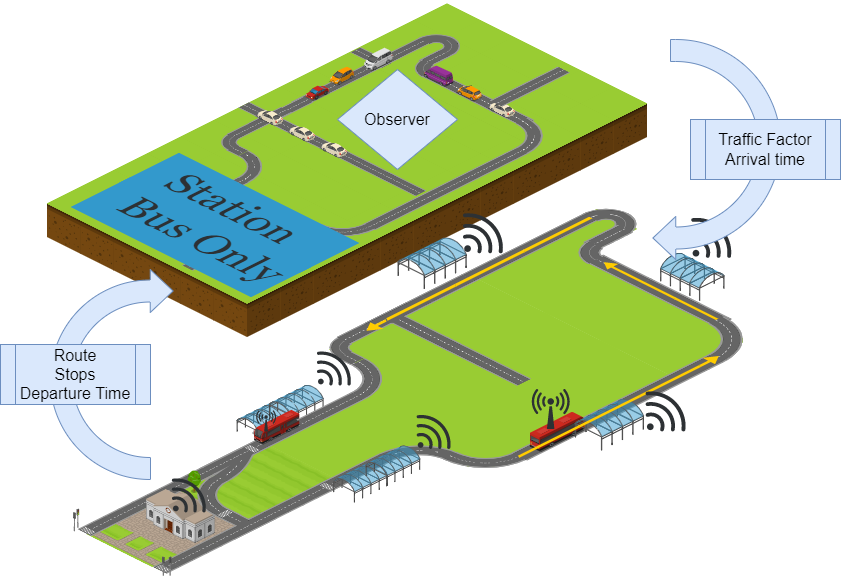}
    \caption{Upper-level observer that estimates horizon T, considered for the solution approach.}
    \label{fig:uplev}
\end{figure}

Upon careful examination of the optimization problem formulated in Problem \ref{problem1}, we recognize that the pure formulation leads to an infinite horizon approach. This is because deriving the optimal policy is contingent on reaching a desired state rather than a fixed time horizon. Considering the complexity of the formulation presented in Section \ref{sec:ModelingFramework}, we propose simplifying the problem by introducing a solution dependent on a fixed time horizon.
We introduce an upper-level observer to implement this approach, as shown in Fig. \ref{fig:uplev}. In our approach, the upper-level observer calculates the time horizon, corresponding to each bus round trip time. The observer knows the traffic factor at each point in time, and the computation is performed after receiving information about the route, number of stops, and departure time of each bus. This approximation is crucial for problem resolution. Before each trip begins, we can estimate the duration of the round trip, providing us with the time horizon $k = T$ at which the bus will return to the depot. With this information, the problem becomes a finite horizon additive cost problem, where the final cost is known. By adopting this fixed-time horizon, we simplify the problem and transform it into a finite horizon optimization task with a known final cost.
\vspace{-0.4cm}
\begin{assumption}
The time horizon $T$ is computed by an upper-level observer before each trip, considering the traffic factor and bus route. The fixed $T$ ensures reaching the desired termination state, accounting for effective planning. 
\end{assumption}
Assumption 1 enables us to formulate the problem as a finite horizon additive cost problem, allowing for the use of dynamic programming to derive the optimal policy. While the assumption can be relaxed for an infinite horizon approach, our focus in this work is to demonstrate the efficacy of the proposed approach in maximizing passenger safety. Relaxation of the assumption can be explored for large-scale implementation and optimization of computational time, which are potential areas for future work. With the horizon T, the final cost is given by $J_T(X_T, U_T) = 0$, and $x_T$ and $u_T$ are known. $x_T$ is the desired terminal state $x_T = (p_T, 0, b_T, m_T)$, where $p_T$ and $m_T$ have fixed finite values given by the specific loop considered, while $u_T = 0$ is the only control action possible after arriving at the depot. From the final state, a typical dynamic programming approach is proposed and solved with a rolling horizon control approximation, referred to, from now on, as the look-ahead (\emph{LA}) horizon.

\section{Simulation and Discussion} \label{sec:Simulation}
\subsection{Simulation Setup}
The system timetable is predetermined based on a frequency-based approach, following the design principles of BRT systems. For our simulations, we used the BELLAIRE TX METRO as a reference and constructed a timetable with four stops, including the depot. The round trip time for the loop is approximately 9-10 minutes, assuming a bus speed of 5 m/s under normal traffic conditions. The standard traffic factor used in the simulations is $c = 5$. The timetable planning includes predefined dwell times for each stop and serves as the baseline for our analysis. The total length of the loop is approximately 1.5 miles (2.4 km), with a distance of 0.373 miles (600 m) between each stop. The length of the stops is 0.0062 miles (10 m), with a corresponding flexibility factor $\delta = 5$. To account for the stochastic nature of the problem, we utilized historical data from \cite{filabadi2022new} and fitted lognormal distributions. The sets $\mathcal{A}={1, 2, 3, 4, 5}$ and $\mathcal{G}={1, 2, 3, 4, 5}$ have independent lognormal distributions, where $\probP(0< A_k\leq a)$ and $\probP(0< G_k \leq g)$ represent the probabilities of the corresponding elements in the sets. The probabilities for $\mathcal{A}$ are $(0.51,0.77,0.88,0.95,1)$, and for $\mathcal{G}$, they are $(0.14,0.81,0.97,0.99,1)$. In our implementation, we evaluate the probabilities associated with the events defined in the finite sets. According to the data, new passengers arrive every 60 seconds, and passenger alighting occurs only at each stop.

\subsection{Simulation Results and Discussion}
To solve Problem 1 using the dynamic solution approach described in Section \ref{sec:control_framework}, we implemented various look-ahead horizons ranging from $4 s$ up to $9 s$. The purpose was to analyze their computational time and average cost performance. We conducted 20 simulations for each look-ahead horizon to compare the results. Using the objective function, we derived an optimal control strategy that determines the speed of the bus, which directly affects its position. The baseline scenario was derived from the timetable to adhere to it closely. Our primary goal is to maximize passenger safety by minimizing their waiting times at stops while keeping the bus speed within an acceptable predefined range. Thus, we aim to arrive at the stops quickly while respecting the imposed speed bounds.

\begin{figure}[ht!]
    \centering
    \includegraphics[trim={4.35cm 0 5.05cm 0},clip, width=0.9 \linewidth]{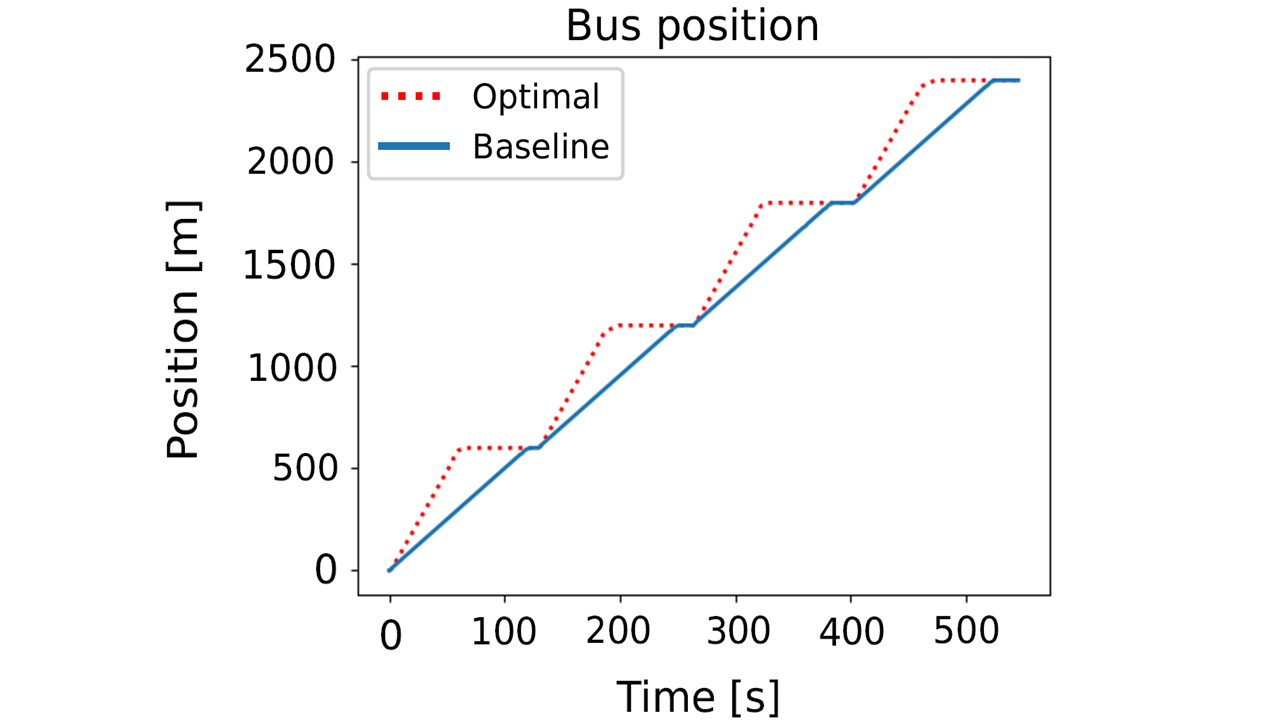}
    \caption{Bus position when applying optimal policy, compared with the baseline.}
    \label{fig:posbase}
\end{figure}

Fig. \ref{fig:posbase} displays the results obtained for the bus position when implementing an optimal policy derived with a $5 s$ look-ahead policy. The bus position achieved with the optimal policy addresses a significant issue of the baseline scenario: the short dwell time at each stop. Although the timetable suggests different dwell times for each stop in the baseline scenario, these times may be short relative to the number of passengers waiting to board.
The optimal policy with longer dwell times improves upon the baseline scenario by addressing the issue of short dwell times. This is beneficial because shorter dwell times can lead passengers to wait longer on the street and have less time to board the bus. The longer dwell times in the optimal policy are not directly determined by the number of passengers but rather by the timetable's departure time constraint.

\begin{figure}[ht!]
    \centering
    \includegraphics[trim={0.3cm 0 2cm 0},clip, width=0.99 \linewidth]{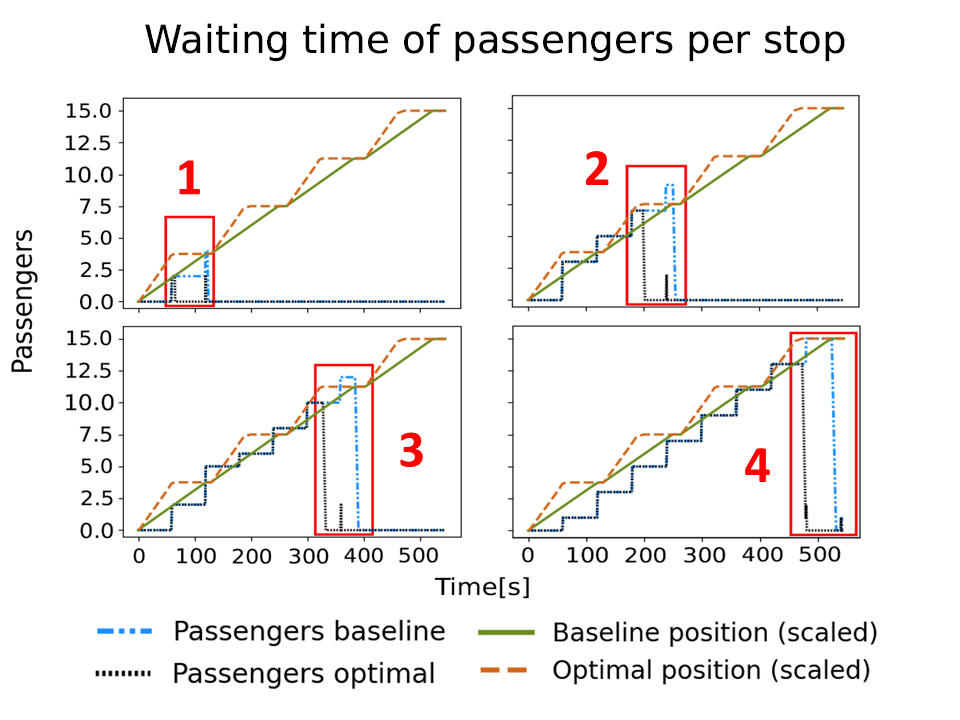}
    \caption{Waiting time of passengers  (optimal policy vs baseline scenario).}
    \label{fig:wtoptvsbl}
\end{figure}

In Fig. \ref{fig:wtoptvsbl}, the waiting time of passengers at each stop is compared between the optimal policy (represented by the black dotted line) and the baseline scenario (represented by the blue dash-dotted line). The red rectangles indicate the corresponding stop numbers, and the position of the bus is scaled and plotted alongside the waiting time for reference. The optimal policy, represented by the orange dashed line, minimizes the waiting time of passengers. This is achieved when the bus arrives earlier at the stop and waits until the timetable allows departure. When passengers arrive while the bus is at the stop, they can board immediately without any additional wait time. This approach reduces waiting time for passengers compared to the baseline scenario.

We also evaluated the performance of our approach  by computing the area under the curve for the passengers' waiting time, and the results indicate the optimal scenario improves the waiting time by at least $21\%$ for the fourth stop and at most $89\%$ for the first stop.  The proposed method was implemented with 6 different look-ahead horizons, from $4 s$ to $9 s$, each look-ahead with 20 runs to compare results. We tracked the control action's average and maximum computing time at each stage $k$. All of the times obtained are in the order of $ms$, specifically less than $10 ms$, which means the proposed framework is implementable in real-time, even if the complexity of the problem is augmented. The shortest average time was obtained at $4 s$ LA, while the shortest maximum time was at $5 s$ LA. Then, we conducted the simulations for the final results with a $5 s$ LA policy, obtaining satisfactory results.

\section{Conclusion} \label{sec:conclusion}

In this work, we proposed an optimal planning framework to maximize passenger safety in a BRT system by minimizing their initial waiting time. Our framework formulates an optimization problem to derive a speed-optimal policy based on the timetable, ensuring \emph{on-time} performance. Through simulations, we demonstrated the effectiveness of our approach, which tracks bus position and maintains a range of desired speeds. Our real-time implementable framework represents a novel approach to minimizing passenger initial waiting time using precise position and speed tracking. Future research can explore large-scale simulations, introduce more stochasticity through traffic factor manipulation, consider infinite horizon approaches, and employ different optimization algorithms. Extensions could include autonomous buses, coordination strategies, and integration with driver assistance systems. Overall, our work contributes to BRT system optimization and offers directions for further research.

\balance

\bibliographystyle{IEEEtran}
\bibliography{references/ids_lab,references/vsche}

\end{document}